\documentstyle[12pt]{article}

\begin{document}

\title{On solving Schwinger-Dyson equations
for non-Abelian gauge theory}
\author{V.E. Rochev\footnotemark \\ 
{\it Institute for High Energy Physics}\\
{\it 142284 Protvino, Moscow region, Russia}}
\footnotetext{rochev@mx.ihep.su}
\maketitle

\abstract{A method for solving Schwinger-Dyson equations for the Green
function generating functional of non-Abelian gauge theory is proposed.
The method is based on an approximation of Schwinger-Dyson equations
by  exactly soluble equations. For the $SU(2)$ model the
first step equations of the iteration scheme are solved which define
a gauge field propagator. Apart from  the usual perturbative solution,
a non-perturbative solution  is found,  which corresponds to the spontaneous
symmetry breaking and obeys infrared finite behaviour of
the propagator.}

\bigskip

\newcommand{\be}{\begin{equation}}
\newcommand{\ee}{\end{equation}}
\newcommand{\ba}{\begin{eqnarray}}
\newcommand{\ea}{\end{eqnarray}}
\section{ Introduction}
 The Schwinger-Dyson equations (SDE) method is one of the basic tools
for investigating of the Green functions of the quantum theory.
Hitherto the one universal method for solving SDEs  has been the coupling
constant perturbation theory (hereafter simply termed perturbation theory).
The range of applicability of other methods (for example, $1/N$-expansion)
is limited by a narrow class of models. In particular, the $1/N$-expansion
method cannot be used in the calculations for 
 non-Abelian gauge theories
due to the complicated structure of the leading approximation.

On the other hand, the applicability of perturbation theory to
the investigation of non-Abelian gauge theories is limited by a deep-Euclidean
region. In the non-perturbative region of small momenta the physical vacuum
of non-Abelian gauge theories obeys the nontrivial structure that is beyond
the framework of perturbation theory.
In the SDE terms this fact can be understood if one takes into account
the radical difference in the properties of the 
leading-approximation equations of the perturbation theory and the original
exact equations. The SDEs for the generating functional of Green functions
are equations in functional derivatives. The leading approximation of
perturbation theory comprises  neglecting  the higher
derivatives terms in these equations 
(just such terms correspond to an interaction).
The leading-approximation equations of perturbation theory are of a lower
order compared with the exact ones;
 therefore, the class of solutions thus described contracts drastically, and
non-perturbative solutions which correspond to the nontrivial physical vacuum
practically drop out of the consideration.
 This feature of SDEs in the non-perturbative
region is noted repeatedly in  simple models
(see \cite{1}). 

In this work a method for the  SDE solution  of non-Abelian gauge theory
is proposed. It takes into account the terms with higher derivatives
(i.e. self-interaction of the non-Abelian fields) {\it ab ovo} in the leading
approximation.
 Though we limit ourselves to the simplest extension of the
class of SDE solutions, the results  are non-trivial:
the non-perturbative solution  which corresponds to  spontaneous
symmetry breaking (with non-Higgs mechanism) is found. It obeys a non-singular
 behaviour of the propagator in the non-perturbative 
infrared region of small momenta.

An idea of the method comprises the approximation of SDEs for the
generating functional by  equations with "constant" (i.e. independent of
sources) coefficients. These equations have a simple exponential solution,
 which is a foundation for
the linear iteration scheme. The method is  universal, as is perturbation
theory, i.e. it is applicable to practically any model of quantum
field theory.  For the scalar $\phi^4$  theory it has been shown in  \cite{2}
 that the method describes  such non-perturbative
phenomena as  spontaneous symmetry breaking and 
the trivialization of  $\phi^4$ theory
at $d=4$. This method has also been successfully applied in the
investigation of the Gross-Neveu model for finite $N$  \cite{3}.
\section{ Schwinger-Dyson equations and the iteration scheme}
 A system of SDEs for the generating functional
$G(J,\eta)$ of the Green functions of non-Abelian gauge theory has the form
\be
{\cal D}^{ab}_{\nu}(\frac{\delta}{i\delta J}) 
F^b_{\nu\mu}(\frac{\delta}{i\delta J}) G +
 \frac{1}{\alpha}\partial_{\mu}\partial_{\nu} 
\frac{\delta G}{i\delta J^a_{\nu}}
+ g f^{abc}\frac{\delta}{\delta \bar \eta^c}
 \partial_{\mu}\frac{\delta G}{\delta \eta^b} + J^a_{\mu} G = 0,
\ee
\be
i\partial_{\mu} {\cal D}^{ab}_{\mu}(\frac{\delta}{i\delta J})
\frac{\delta G}{\delta \bar \eta^b} + \eta^a G = 0.
\ee
Here $F^a_{\mu\nu}(A) = \partial_\mu A^a_\nu - \partial_\nu A^a_\mu +
g f^{abc} A^b_\mu A^c_\nu$ is the gauge field tensor,
${\cal D}^{ab}_{\mu}(A) = \delta^{ab}\partial_{\mu} -
g f^{abc}A^c_{\mu}$ is the covariant derivative, 
 $f^{abc}$ are structure constants of a gauge group,
 $J^a_\mu (x)$ is the source of the gauge field,
$\eta^a(x)$ is the source of a ghost field, $\alpha$ is
the gauge parameter, $g$ is the coupling constant.
We work in  Minkowski space with a metric 
(1, -1, -1, -1), and 
$x_\mu y_\mu \equiv g_{\mu\nu}x^\mu y^\nu$ by definition.

At $g=0$  the system (1)-(2) has a solution
$$G_{free} = \exp\{ \frac{1}{2i} J \ast D_{free} \ast J + 
i \bar\eta \ast \Delta_{free} \ast \eta\}, $$
where $D_{free}$ and $\Delta_{free}$ are free propagators of the gauge
field and ghost field respectively, and 
$$J\ast D_{free}\ast J\equiv
\int dx dy J^a_\mu(x)(D_{free})^{ab}_{\mu\nu}(x-y)J^a_\nu(y),$$ etc.
The functional $G_{free}$ is 
the generating functional of the free Green functions.
The iteration scheme based on
$G_{free}$    is the usual coupling constant
perturbation theory.

We shall use an alternative iteration scheme for SDEs (1)-(2), which
is formulated as follows: the leading approximation
is a system of equations with terms proportional to the sources
 $J$ and $\eta$ omitted. (For the system of eqs. (1) and (2) these are
the last terms.) This system has a  solution
$$G_0 = \exp i \{ J\ast V + \bar\eta \ast C + \bar C \ast \eta \},$$
where $J~\ast~V~\equiv~\int~d~x~J^a_\mu(x)~V^a_\mu(x)$, etc.
The coefficient functions $V^a_\mu$ and $C^a$ are solutions of
the corresponding system of "characteristic equations".
When constructing the iteration scheme for the generating functional
 $$G = G_0 + G_1 +\cdots + G_n +\cdots$$
 the omitted terms $JG$ and $\eta G$ should be considered
as perturbations, i.e. equations of the iteration scheme are
\be
\{ {\cal D}_{\nu}(\frac{\delta}{i\delta J}) 
F_{\nu\mu}(\frac{\delta}{i\delta J})  +
 \frac{1}{\alpha}\partial_{\mu}\partial_{\nu} 
\frac{\delta }{i\delta J_{\nu}}
+ g f\frac{\delta}{\delta \bar\eta}
 \partial_{\mu}\frac{\delta }{\delta \eta}\}G_n = - J_{\mu} G_{n-1},
\ee
\be
i\partial_{\mu} {\cal D}_{\mu}(\frac{\delta}{i\delta J})
\frac{\delta G_n}{\delta \bar\eta} =- \eta G_{n-1}.
\ee

A solution of eqs.(3)-(4) has the form $G_n = P_n G_0$,
where $P_n$ is a polynomial in $J$ and $\eta$.
Therefore, at  each iteration  step  we obtain a closed system of
equations for coefficient functions of the polynomial $P_n$, which
completely defines the Green functions of the given step. There is no
 manifest small parameter in the usual sense in this scheme: "smallness"
is defined by the condition that 
the Green functions are  derivatives of the generating functional
at $J=\eta=0$, and it is sufficient for us to know
$G(J,\eta)$ near zero, i.e. in the region where the neglected terms are small. 
At each step of the iteration we approximate the functional
 $G/G_0$ by a sum of the polynomials
 $P_n$, and a degree of the polynomial increases with  each step.
As is known, for ordinary differential equations the scheme of this
type is equivalent to  iterations of Volterra-type integral equations
and gives an expansion that converge well. That is why we may hope that 
this scheme possesses  good convergence properties.
 In any case it is clear that the convergence of this scheme
is no worse than that of the perturbation theory. The perturbation
theory is singular in the sense of  differential equation theory, since
the higher derivatives are omitted in the leading approximation.
In contrast to the singular perturbation theory, the scheme proposed
is regular in the sense above. This circumstance gives us  hope 
for improving the convergence properties (see \cite{4} for more
discussion).

To remove the ultraviolet divergences   it is necessary to 
supplement SDEs (1)-(2) and the iteration scheme equations (3)-(4)
with the corresponding counterterms. The counterterms  are also defined by the 
iteration procedure:
 $\delta z=\delta z_0+\delta z_1+\cdots$, i.e. 
at each step it is necessary 
to take into account the counterterms of the corresponding
order.
 
Let us consider the leading approximation in more detail. As 
has been noted above
the solution of the leading approximation equations is the linear exponential
in the sources. The vacuum structure is defined by the solutions of 
the characteristic equations for $V^a_\mu$ and $C^a$
\ba
{\cal D}_\nu (V) F_{\nu\mu} (V) + 
\frac{1}{\alpha} \partial_\mu \partial_\nu V_\nu + gfC\partial_\mu\bar C
= 0,\\
\partial_{\mu} {\cal D}_\mu (V) C = 0 .\nonumber
\ea
These equations have a great number of solutions which reflects
the non-trivial vacuum structure of the non-Abelian gauge theory.
In this connection it is appropriate to recall 't Hooft's
conjecture \cite{5} on the existence of different vacuum modes
for non-Abelian gauge theory at zero temperature:
a superconducting one (spontaneously broken), dual superconducting
(confinement), etc. Realization of either mode depends on the values
of some quantities (manager parameters) whose definition is a problem
of the quantum field dynamics.  It is 
reasonable to suppose that the whole set of solutions of characteristic
equations (5) is divided into  classes, each corresponds
to some mode of the gauge theory in the above sense .

In this paper we limit ourselves  to  investigation of the
first step of the iteration scheme based on the simplest
solutions of the characteristic equations (5).

The trivial solution of eqs.(5) $V^a_\mu = C^a = 0$ (i.e., $G_0 = 1$) leads
 to an iteration scheme which is simply the
reconstructed perturbation theory. 
The simplest extension of the set of vacuum solutions compared
with the trivial perturbative vacuum comprises taking into account
non-zero constant solutions of eqs.(5).
 Namely, we choose the leading approximation in the form
\be
G_0 = \exp i\{J\ast V\},
\ee
where the vacuum vectors $V^a_\mu$ are independent
of the space-time variables and satisfy the condition
\be
f^{abc}f^{cdh} V^b_\nu V^d_\nu V^h_\mu = 0,
\ee
which follows from the characteristic equations (5), if $C= \partial V = 0$.
Below we shall consider only this class of the leading approximation solutions.
Surely we have no prior physical foundation (except for reasons of simplicity) 
 for the choice of this class of solutions. Nevertheless,
we shall see that this simple set of solutions leads to a
new spontaneously broken mode.

Concluding our discussion of the leading approximation, let us touch upon 
the definition of the ground state ("physical vacuum"). Due to the linearity
of SDEs (1)-(2) and the iteration scheme equations  (3)-(4) an arbitrary
linear combination of solutions is also a solution. Therefore,
the physical vacuum functional $G=<0\mid 0>_J$ should be constructed
as a linear combination of partial solutions, each of them corresponding
to some solution of the characteristic equations (5). 
In other words, the physical vacuum is a superposition of $V$-vacua.
This combination  should be chosen in such a way as to maintain the
admissible physical properties of the Green functions (for example,
$<0\mid A^a_\mu\mid 0> = 0$) and the energy minimality condition.

To formulate the iteration scheme in terms of the polynomials $P_n$,
it is convenient to introduce the matrix quantity
\be
W^{ab}_\mu = ig f^{abc} V^c_\mu.
\ee
Then, the iteration scheme equations have the form of equations for
the polynomials $P_n$:
\ba
\{ [ {\cal D}_\nu (V) + igf\frac{\delta}{\delta J_\nu} ] 
[ F_{\nu\mu} (\frac{\delta}{i\delta J}) + W_\nu \frac{\delta}{\delta J_\mu} -
W_\mu \frac{\delta}{\delta J_\nu} ] + \nonumber \\ 
i [ W_\mu , W_\nu ] \frac{\delta}{\delta J_\nu}
+ \frac{1}{\alpha} \partial_\mu \partial_\nu \frac{\delta}{i\delta J_\nu} +
g f \frac{\delta}{\delta \bar\eta}\partial_\mu \frac{\delta}{\delta \eta}\}P_n
= - J_\mu P_{n-1},
\ea
\ba
\partial_\mu [ {\cal D}_\mu (V) + 
ig f \frac{\delta}{\delta J_\mu}]\frac{\delta P_n}{\delta \bar\eta} =
i\eta P_{n-1}.
\ea
Here $P_0$ = 1. 
The first-step solution $P_1$ 
 defines the leading approximation for the propagators of the gauge and
ghost fields. The following steps define the  many-particle functions.
\section{Solution of  first-step equations}
The solution of  first-step equations of the iteration scheme (9)-(10) is
the quadratic polynomial in the sources 
\be
P_1 = \frac{1}{2i} J\ast D\ast J + i\bar\eta\ast\Delta\ast\eta.
\ee
Eqs. (9) and (10) give us equations for the functions
$D^{ab}_{\mu\nu}(x-y)$ and
$\Delta^{ab}(x-y)$. The equation for $D_{\mu\nu}$ 
can be essentially simplified  by modifying the gauge condition.
Instead of the usual covariant gauge  
used above, it is convenient to use the modified gauge condition
("$V$-gauge") with the gauge fixing term 
\be
{\cal L}_{gauge}=-\frac{1}{2\alpha} ({\cal D}_\mu(V)A_\mu)^2.
\ee
The ghost terms should be changed correspondingly. For the transition
to the $V$-gauge in formulae (1)-(4), (5) and (9)-(10),
it is sufficient to perform the substitution 
\be
\partial_\mu \rightarrow {\cal D}_\mu (V).
\ee
An essential circumstance is that
 the leading approximation condition (7) is not changed in the case.

The equations have a particularly simple form in the gauge
 $\alpha=1$ ("diagonal $V$-gauge"). Then, the equation for $D_{\mu\nu}$
in momentum space is
\be
\{ {\cal K}^2 g_{\mu\nu} + 2 [{\cal K_\mu , K_\nu}]\} \tilde D_{\nu\lambda}(k)
= - g_{\mu\lambda},
\ee
where $k$ is the momentum, and the following notation has been introduced:
\be
{\cal K}^{ab}_\mu = k_\mu \delta^{ab} - W^{ab}_\mu.
\ee
In the region of large $k$ eq. (14) tends to the equation for
the free propagator in the diagonal gauge, i.e. at $k\rightarrow\infty$
\be
\tilde D_{\mu\nu} (k) \approx -\frac{1}{k^2} g_{\mu\nu}.
\ee
The equation for the ghost propagator $\Delta$ is
\be
{\cal K}^2 \tilde\Delta (k) = -1.
\ee
In the large $k$ region the propagator $\tilde\Delta$ 
also tends to the free propagator.
Therefore, the ultraviolet behaviour of the solutions with a nontrivial
vacuum vector ${\bf V}_\mu$ is the same as for the usual perturbation theory.

Below we consider the case of $SU(2)$ gauge group
and restrict ourselves to the subset of constant ${\bf V}_\mu$
with zero field tensor $F_{\mu\nu}(V)=0$, or, equivalently, with
\be
\epsilon^{abc} V^b_\mu V^c_\nu = 0.
\ee
It is clear that  condition (18) ensures the leading approximation
condition (7). The energy of that kind of solutions is zero, as for
the perturbative solution with trivial vacuum. In that sense this set
of solutions can be termed  {\it quasiperturbative}.
Eq. (18) gives us
 $[{\cal K_\mu , K_\nu}] = [W_\mu , W_\nu] = 0$, and the solution of
 eq.(14) is reduced to the inversion of  the matrix ${\cal K}^2$.
The result is
\be
\tilde D^{ab}_{\mu\nu} (k) = -g_{\mu\nu} [ \frac{p}{p_1} \delta^{ab} +
\frac{2(kW^{ab})}{p_1} + (\frac{1}{k^2} - \frac{p}{p_1}) \frac{(V^aV^b)}{V^2}],
\ee
where the following notations has been introduced:
\be
p(k,V) = k^2 + g^2V^2, \; p_1(k,V) = p^2 - 4g^2(kV)^2.
\ee

A solution of the ghost propagator equation (17)  is also given by formula (19)
(without  $g_{\mu\nu}$).

Except for eqs. (14) and (17), the first-step equations give one more relation
that contains a quantity
 $D_{\mu\nu}(0)$, which should be understood as some regularization.
In essence this relation is a condition for the first step counterterms
 $\delta z_1$.
 (There is no need to introduce the leading approximation
counterterms in this case, i.e. $\delta z_0 = 0$).
Since the Green functions of the first step are finite, this condition
for the counterterms is necessary to remove the ultraviolet divergences  
at {\it the second step} of the scheme.
This peculiarity of the given iteration scheme is displayed here in exactly
the same manner as for the scalar field theory (see \cite{2}).
\section{Gauge field propagator}
 Let us turn now to a possible physical interpretation of the solutions.
At ${\bf V}_\mu = 0\; D_{\mu\nu}$ and $\Delta$ are free propagators
of gauge and ghost field, and  the whole iteration scheme is a reconstructed
series of the perturbation theory.
At ${\bf V}_\mu \neq 0$ the situation is more complicated. It is clear that
in this case it is difficult to interpret the function 
 $D_{\mu\nu}$ given by (19) as the propagator of a particle in the
Poincar\'e invariant theory.
Let us recall, however, that we have a number $\{G(V)\}$ of  solutions
of SDEs, each corresponding to some vector ${\bf V}_\mu$ satisfying
 the condition (18).
Due to eq.(18) the field strength tensor is identically zero and any such
solution possesses  zero energy. Therefore, a candidate for the
 "physical vacuum functional"
for this set of solutions is a superposition of all solutions.
We shall exploit this fact in the construction 
of a Poincar\'e invariant solution
which can be interpreted as a particle propagator, i.e. a function depending
only on the momentum $k$ and the scalar quantity
\be
v^2 = V^2  \equiv V^a_\mu V^a_\mu.
\ee
(The quantity $v^2$ plays the role of an order parameter.) The 
construction  is equivalent, in essence,
to some averaging, i.e. an integration with a measure 
 $d\mu(V)$, we shall therefore denote it by 
angular brackets: $<G> = \sum_V G(V)$, etc.
In the foundation 
of this operation we set the following
conditions
\be
<V^a_\mu> = 0,\,\,\,\,\,<V^2> = v^2.
\ee
Their necessity for the Poincar\'e invariant theory is evident.
It is  also evident  that
\be
<V^a_\mu V^b_\nu> = \frac{1}{4} v^2 g_{\mu\nu} E^{ab},
\ee
where $tr\, E = 1$. To determine the form of the matrix
 $E^{ab}$, consider the leading approximation condition (18).
The geometrical meaning of  condition (18) is the 
collinearity of the vectors
 ${\bf V}_\mu$ in isotopic space. Consequently, at
 ${\bf V}_\mu \neq 0$ there exists a selected direction in 
isotopic space. This direction can be chosen as a basis vector, for
example ${\bf n}_3$.
In this basis $V^a_\mu = \delta^{a3} v_\mu$, and 
$E^{ab} = \delta^{a3} \delta^{b3}$.
Therefore, at ${\bf V}_\mu \neq 0$ the isotopic symmetry is spontaneously
broken.

Further calculation is reduced to the averaging of the functions
 $f((kV)^2) = p/p_1$ and $f^{ab} = f\cdot (V^a V^b)/V^2$.
First of all, note that at $k~\rightarrow~0\; f\rightarrow 1/g^2v^2$,
and at $k\rightarrow\infty\;f\rightarrow 1/k^2$. These Poincar\'e-invariant
properties of $f$ should, of course, be conserved for
 $<f>$ as well.

With eqs.(22)-(24) and its generalizations for an arbitrary monomial in
 ${\bf V}_\mu$,
the following formulae can be proved
\ba
<(kV)^{2n}> = (k^2 v^2)^n \frac{\Gamma (n + 1/2)}{(n+1)! \Gamma (1/2)},\\
<(kV)^{2n}(V^a V^b)> = E^{ab} v^2 <(kV)^{2n}>, \nonumber
\ea 
which are necessary for the calculation of
 $<D_{\mu\nu}>$. The result of the calculation in the above basis is
as follows:
\be
<\tilde D^{33}_{\mu\nu}(k)> = - g_{\mu\nu} \frac{1}{k^2}, 
\ee
\be
<\tilde D^{11}_{\mu\nu}(k)> = <\tilde D^{22}_{\mu\nu}(k)> =
- g_{\mu\nu} \frac{k^2+g^2v^2}{2g^2v^2k^2}\Biggl(1-
\sqrt{1-\frac{4g^2v^2k^2}{(k^2+g^2v^2)^2}}\Biggr).
\ee
(Other isotopic components are equal to zero.)
Therefore, along the selected isotopic direction the particle
propagates as a free one, but along other directions 
a separation of the region of momenta exists, the scale of the
separation being the quantity $g^2v^2$.
It is necessary to stress that both the limiting cases  above
($k^2\rightarrow 0$ and $k^2\rightarrow \infty$) belong to the
region of applicability of the calculations performed, which
is defined by the condition $\mid 4g^2v^2k^2/(k^2+g^2v^2)^2\mid < 1$.
Consider the question about  analytical continuation.
Eq. (26) defines two analytical functions depending on the choice of
 branch of the function $\sqrt{z^2}$, but neither  satisfies
  both the  above asymptotical conditions simultaneously
 and, consequently, these functions are not  solutions of the problem.  
Hence, as a solution one should choose  at 
large $k$  the branch
with the behaviour $1/k^2$, and at small $k$ - another branch which
is a constant $1/g^2v^2$.
At the points $\pm g^2v^2$ the solution goes from one branch to another,
i.e. from the "perturbative" sheet  to the "non-perturbative" one.
At the point $k^2=g^2v^2$ the solution is continuous, and at the point
 $k^2=-g^2v^2$ has a discontinuity. (Note that the first-step calculations of
the iteration scheme do not fix the sign of $v^2$.)
Such unusual features of the solution near the separation points 
 are  likely to be connected  with our limitation 
of the characteristic equation solutions. Probably an extension of the class of
 solutions will lead to smoothing of the propagator behaviour near
the  points of  separation of the perturbative and non-perturbative regions.
\section{Discussion}
In this paper we have attempted to introduce a new iteration scheme
to solve the Schwinger-Dyson equations for non-Abelian gauge theory.
As a very first step we have considered a scheme based on the 
simplest constant solutions of the characteristic equations (5).
Of course, there are no prior grounds for expecting such a set of solutions
to be useful for the physical applications. Some
other sets of more non-trivial solutions (instanton like and so on)
of the characteristic equations are likely to be of real physical
interest. Dealing with  coordinate-dependent solutions 
is a more difficult task;
in particular, the averaging procedure of Sec.4 should be essentially
modified.
Another important question  is the correspondence
 of the proposed method 
with the functional integral approaches, such as well-known
background field method \cite{6}. In spite of their evident similarity,
the proposed
iteration scheme seems to be somewhat different: the 
 structure of expansions is distinct.
The problem of the correspondence 
 requires further investigations based
on the calculation of higher steps of the iteration scheme.

The set of quasiperturbative solutions of Sects. 3-4 leads to
spontaneous breaking of $SU(2)$ gauge symmetry. The question
arises as to whether such solutions can be used to construct a realistic
model of electroweak interactions in the spirit of dynamical
gauge symmetry breaking models without Higgs bosons (see \cite{7}
and refs. therein). Further investigation which should
essentially extend the class of solutions is necessary to eliminate
this question. On the other hand, the set of quasiperturbative
solutions satisfying  condition (18) considered here does not exhaust
 the solutions with zero energy, and some further
averaging to restore the symmetry may be necessary. 

For $SU(3)$ gauge group (the case of QCD) the variety of solutions
 increases greatly in comparison with $SU(2)$
group (see, for example, \cite{8}) .
Avoiding  discussion of this case, nevertheless note
 that the infrared finite
behaviour of the gluon field propagator 
 has recently been discussed in detail  (see 
\cite{9} and refs. therein).
\section*{Aknowlegements}
The author is grateful to A.I.~Alekseev, B.A.~Arbuzov, P.A.~Saponov
and V.V.~Vladimirsky for useful discussions.
 The work is supported by RFBR, grant No.95-02-03704.

\end{document}